\documentclass[12pt]{article}
\usepackage{amsfonts}
\usepackage{amssymb}
\usepackage{tikz}
\usepackage{amsmath,amssymb}
\usepackage{bbm}

\def\sfrac#1#2{{\textstyle\frac{#1}{#2}}}

\newcommand{\CU}{\mathcal{U}}    
\newcommand{\CL}{\mathcal{L}}    
   
\newcommand{\CE}{\mathcal{E}}  

\newcommand{\CH}{\mathcal{H}}
\newcommand{\CJ}{\mathcal{J}}
 
\newcommand{\CO}{\mathcal{O}} 
\newcommand{\Z}{\mathbb{Z}}   
   
\newcommand{\R}{\mathbb{R}}     
\newcommand{\C}{\mathbb{C}}     
\newcommand{\Nbb}{{\mathbb{N}}}    

\newcommand{\im}{\mathrm{i}} 
 
\newcommand{\dd}{\mathrm{d}}     
\newcommand{\dpar}{\partial}     
     
\newcommand{\+}{\dagger}

\newcommand{\sU}{\mathrm{U}}

\newcommand{\sSO}{\mathrm{SO}}     
\newcommand{\sGL}{\mathrm{GL}}

\newcommand{\al}{{{\alpha}}} 
  
\newcommand{\dal}{{{\dot\alpha}}}

\newcommand{\veps}{{\varepsilon}} 
\newcommand{\vph}{{{\varphi}}}

\newcommand{\xh}{\hat{x}} 
\newcommand{\ph}{\hat{p}}
\newcommand{\zh}{{\hat{z}}}

\newcommand{\zb}{{\bar{z}}}

\newcommand{\Av}{{A_{\sf{vac}}}}

\newcommand{\Jv}{{J_{\sf{v}}}}

\newcommand{\Fv}{{F_{\sf{vac}}}}
\newcommand{\Qv}{{Q_{\sf{v}}}}
\newcommand{\Uv}{\sU(1)_{\sf{v}}}

\newcommand{\und}{{\quad\mathrm{and}\quad}}
\newcommand{\with}{{\quad\mathrm{with}\quad}}
\newcommand{\for}{{\quad \mathrm{for}\quad }}

\makeatletter

\@addtoreset{equation}{section}
\makeatother

\begin{document}
\begin{titlepage}
\setcounter{page}{0}
.
\vskip 20mm
\begin{center}
{\Large \bf Quantum Charge as a Particle{/}Antiparticle Label}\\
\vskip 1.2cm
{\large Alexander D. Popov}
\vskip 0.7cm
{\em Institut f\"{u}r Theoretische Physik,
Leibniz Universit\"{a}t Hannover\\
Appelstra{\ss}e 2, 30167 Hannover, Germany}\\
{Email: alexander.popov@itp.uni-hannover.de}
\vskip 1cm
\end{center}
\begin{center}
{\bf Abstract}
\end{center}
We show that the concept of particles and antiparticles is related to the Weyl-Heisenberg group $W\!H_3$. Its generators are the position operators $\hat x^a$, the momentum operators $\hat p_a$, and the central element $Q_{\sf{v}}$, which defines a subgroup U(1)$_{\sf v}$ of $W\!H_3$. We show that the wave functions have a charge relative to the group U(1)$_{\sf{v}}$ equal to either $q_+=1/\hbar$ (particle $\psi_+$) or $q_-=-1/\hbar$ (antiparticle $\psi_-$). We call these eigenvalues of the operator $Q_{\sf{v}}$ quantum charges: $Q_{\sf{v}}\psi_\pm^{}=q_\pm^{}\psi_\pm^{}$. The wave functions $\psi_+$ and $\psi_-$ belong to the Hilbert spaces $\CH_+$ and $\CH_-$, on which unitarily inequivalent representations $\pi_\hbar$ and $\pi_{-\hbar}$ of the Weyl-Heisenberg group $W\!H_3$ are defined. These representations are related to each other by the antilinear charge conjugation operator $\hat C$, which maps $\CH_+, q_+$ to $\CH_-, q_-$. Charges $q_\pm^{}$ are intrinsic characteristics of the wave functions $\psi_\pm^{}$, similar to spin and electric charge. The wave function possesses one of these charges, $q_+$ or $q_-$, prior to the specification of any equations. We show that negative probabilities/energies and the Klein paradox in Klein-Gordon and Dirac theories are consequences of the violation of U(1)$_{\sf v}$-invariance in these theories. Restoring this invariance separates particle and antiparticles and eliminates all paradoxes at the level of first-quantized relativistic theories.

\end{titlepage}

\newpage
\setcounter{page}{1}

\section{Introduction and summary}

\noindent{\bf Relativistic equations.} Non-relativistic quantum mechanics considers only particles. Within this framework, antiparticles do not exist, and their introduction is considered impossible. The Schr\"odinger equation describes spinless particles, while the Pauli equation for two-component spinors describes spin-$\sfrac12$ particles. The Klein-Gordon and Dirac equations serve as generalizations of the Schr\"odinger and Pauli equations, respectively. These relativistic equations are constructed for a hypothetical free {\it particle} with a wave function $\psi$ ($\psi\in\C$ or $\psi\in\C^4$), as they are viewed as a generalization of the equations of non-relativistic theory, in which the concept of antiparticles does not exist. Formally, this can be understood as the absence of a label on the wave function $\psi$ that would allow distinguishing between a particle and an antiparticle.

So, in the standard approach, $\psi$ is treated as the wave function of a particle. When constructing solutions to the free Klein-Gordon and Dirac equations, it turns out that these solutions fall into two classes -- $\psi=\psi_++\psi_-\in\CH_+\oplus\CH_-$ -- distinguished by the sign of the frequency with respect to the operator $\im\hbar\,{\dpar_t}$, where $t{=}x^0\!{/}c\in\R$ is the coordinate time. Thus, the wave function $\psi$ acquires a ``$\pm$" label, but only post factum - after the equations are solved, not before. Single-particle quantum mechanics is unable to adequately interpret this frequency label and treats negative modes as an anomalous states of the particle (negative probabilities and energies, the Klein paradox). For this reason, it was proposed to shift to quantum field theory (QFT), where the initial object is an infinite-dimensional space of solutions $\CH = \CH_+\oplus\CH_-$, decomposed according to the sign of the frequency. The subspaces  $\CH_+$ and $\CH_-$ in $\CH$ were associated with particles and antiparticles. In light of the foregoing, the concept of an antiparticle is considered to arise only after solving relativistic equations and to be adequately describable solely within the framework of QFT, specifically through the space of negative-frequency field component. We will show that this standard understanding is not correct.

\noindent{\bf  Quantum charge.} In this paper, we show that in non-relativistic quantum mechanics there exists a parameter -- which we shall call the quantum charge -- that distinguishes particles from antiparticles. Moreover, antiparticles are just as natural and essential as particles. The reason lies in the very foundations of quantum mechanics. The operators of coordinates and momenta, together with the central element, form the  Lie algebra of the Weyl-Heisenberg group $W\!H_3\cong\R^6\times\Uv$, which is a central extension of the translation group of the phase space $\R^6$. The central subgroup of this group is group $\Uv$ with generator $\Qv =\im\dpar_\theta$. The wave functions $\psi_+$ and $\psi_-$ of particle and antiparticle belong to the basis complex conjugate representations of the group $\Uv$, which, under the action of the group $\Uv$, transform according to the rules
\begin{equation}\label{1.1}
\psi_+\mapsto e^{-\im\theta}\psi_+\und \psi_-\mapsto e^{\im\theta}\psi_-\quad \Rightarrow\quad 
\Qv\psi_\pm =q_\pm\psi_\pm =\pm\psi_\pm\ .
\end{equation}
We will call $q_\pm =\pm 1$ quantum charges -- these are the weights of the representations of the group $\Uv\subset W\!H_3$. Upon transition to dimensional coordinates, these charges become dimensional constants $q_\pm =\pm 1/\hbar$, where $\hbar$ is Planck's constant. 

Fixing the quantum charge $q_+$ or $q_-$ implies a choice of the Hilbert space -- $\CH_+$ or $\CH_-$ -- to which the wave function belongs. The charges $q_+$ and $q_-$ correspond to the complex conjugate canonical commutation relations. Unitarily inequivalent irreducible representations $\pi_\hbar^{}$ and $\pi_{-\hbar}^{}$ of the Weyl-Heisenberg group $W\!H_3$ are defined on the spaces $\CH_+$ and $\CH_-$. These representations are related by the antilinear charge conjugation operator $\hat C$. Quantum charge is defined prior to the introduction of any equations, as it is an intrinsic characteristic of the wave function.

\smallskip

\noindent{\bf Complex line bundles $L_\C^\pm$.} The Weyl-Heisenberg group is a principal bundle over the phase space $\R^6$,
\begin{equation}\label{1.2}
W\!H_3\stackrel{\Uv}{\longrightarrow}\R^6\ ,
\end{equation}
with $\Uv$ as the structure group of this bundle $W\!H_3=P(\R^6, \Uv)$. Associated with the principal $\Uv$-bundle \eqref{1.2} are two complex conjugate line bundles $L_\C^+$ and $L_\C^-$, whose sections $\psi_+$ and $\psi_-$ have charges $q_+$ and $q_-$ introduced in \eqref{1.1}. The Hilbert spaces $\CH_+$ and $\CH_-$ of unitarily inequivalent irreducible representations $\pi_\hbar$ and  $\pi_{-\hbar}$ of the group $W\!H_3$ are the spaces of square-integrable polarized sections of the line bundles $L_\C^+$ and $L_\C^-$.

\smallskip

\noindent{\bf Evolution of wave functions.} The fact that the wave functions of particles and antiparticles belong to complex conjugate Hilbert spaces $\CH_+$ and $\CH_-$ implies that $\psi_+$ and $\psi_-$ satisfy the conjugate Schr\"odinger equations for spin $s=0$ and the conjugate Pauli equations for spin $s=\sfrac12$. It follows that the frequency signs in the exponents $\exp(\mp\im\omega t)$ are unrelated to the sign of the energy $E=\hbar\omega$, which is the same for the particle and the antiparticle. In fact, the generators of time translation in spaces $\CH_\pm$ are the operators $\hat E_\pm^{}=\pm\im\hbar\dpar_t$, which in the relativistic case guarantees the positivity of the energy of free antiparticles. In the relativistic case, we introduce the evolution of wave functions in the particles' proper time $\tau$ using six-dimensional trajectory space $X^6$ which is a disjoint union of coadjoint orbits of the Poincare group for free particles and antiparticles. We will show that particles and antiparticles remain distinguished by the quantum charges $q_\pm$ of sections of the bundles $L_\C^\pm$ over $X^6$. The evolution of the wave function $\psi_+$ and $\psi_-$ with respect to the parameter $\tau$ leads to the Klein-Gordon equation for spin $s=0$ and the Dirac equation for spin $s=\sfrac12$, with separate equations obtained for $\psi_+$ and $\psi_-$.

The Klein paradox in standard Klein-Gordon and Dirac theories arises because their Lagrangians are not invariant under transformations of the internal symmetry group $\Uv$. This is due to the fact that the positive- and negative-frequency parts $\psi_+$ and $\psi_-$ of the free field $\psi = \psi_++\psi_-$ carry opposite quantum charges and transform in opposite ways under the action \eqref{1.1} of the group $\Uv$. We will show that transition to $\Uv$-invariant Lagrangians eliminates all paradoxes at the level of first quantization, prior to the transition to QFT. In effect, $\Uv$-invariance leads to the separation of the relativistic equations for particles and antiparticles -- labeled by the quantum charges $q_+$ and $q_-$ -- and to the independent conservation of these charges. This does not mean that one should not introduce an interaction between fields $\psi_+$ and $\psi_-$ at the second-quantized level that violates particle number conservation in the system. However, it needs to be done with an understanding of every step in the construction of the theory.

\section{Heisenberg group $H_3$}

 In this and the following sections, we will use standard definitions from differential geometry and group theory, which can be found e.g. in \cite{KobNom, Sternberg, Kir, Folland}.

\noindent{\bf Group $H_3$.}   The seven-dimensional Heisenberg group $H_3\cong\R^7$ is a central extension of the abelian translation group $\R^6$ by the group $\R$. Its elements can be specified as 5$\times$5 matrices,
\begin{equation}\label{2.1}
g=\begin{pmatrix}1&p&\theta + \sfrac12 p{\cdot}x\\
0&1_3&x\\ 0&0&1\end{pmatrix}\ ,
\end{equation}
where $p=(p_a)$ is a row, $x=(x^a)$ is a column, all coordinates are dimensionless, $p\cdot x=p_ax^a$, and $a=1,2,3$. The multiplication law in group $H_3$ takes the form
\begin{equation}\label{2.2}
(p_1, x_1, \theta_1)\circ (p_2, x_2, \theta_2)=\bigl(p_1+p_2, x_1+x_2, \theta_1+\theta_2+\sfrac12(p_1\cdot x_2-p_2\cdot x_1)\bigr)\ ,
\end{equation}
where the last term demonstrates the non-commutativity of translations following the central extension.

\noindent{\bf One-forms on $H_3$.}   On the Lie group manifold $H_3$, we introduce right-invariant one-forms:
\begin{equation}\label{2.3}
(\dd g)g^{-1}=\begin{pmatrix}0&\dd p&\dd\theta{+}\sfrac12( p{\cdot}\dd x{-}x\cdot\dd p)\\
0&0&\dd x\\ 0&0&0\end{pmatrix}\ ,
\end{equation}
\smallskip
\begin{equation}\label{2.4}
\Rightarrow\ \Theta^p=(\dd p_a),\quad \Theta^x=(\dd x^a)\und \Theta^\theta =\dd\theta+\sfrac12(p_a\dd x^a-x^a\dd p_a)\ .
\end{equation}
From formulae \eqref{2.3}-\eqref{2.4} follow the Maurer-Cartan equations of the form:
\begin{equation}\label{2.5}
\dd\Theta^\theta =\dd p_a\wedge\dd x^a=:\omega\und \dd\Theta^p=\dd\Theta^x=0\ .
\end{equation}
Here, $\omega$ is a symplectic 2-form on the space $T^*\R^3\cong\R^6\subset H_3$ (three coordinates $x^a$ and three momenta $p_a$).

\noindent{\bf Vector fields on $H_3$.} The right-invariant vector fields on the group manifold $H_3$, dual to the one-forms \eqref{2.4}, have the form
\begin{equation}\label{2.6}
\nabla_{x^a}^{}=\frac{\dpar}{\dpar x^a}+\sfrac12 p_a\Jv\ ,\quad 
\nabla_{p_a}^{}=\frac{\dpar}{\dpar p_a}-\sfrac12 x^a\Jv\ ,\quad 
\nabla_{\theta}^{}=\Jv:=-\dpar_\theta\ .
\end{equation}
These vector fields form a basis of the Heisenberg algebra with non-zero commutators of the form
\begin{equation}\label{2.7}
[\nabla_{p_a}^{}, \nabla_{x^b}^{}]=\delta^a_b\Jv\ .
\end{equation}
Here, $\Jv$ is a generator of the center $\R$ of the Heisenberg group $H_3$, and symplectic space $\R^6=H^3/\R$ is a homogeneous space of the group $H_3$.

\section{Weyl-Heisenberg group $W\!H_3$}

\noindent{\bf Group $W\!H_3$.}    In quantum mechanics, instead of $H_3$ one considers the reduced Heisenberg group $W\!H_3:=H_3/\Z$, in which the center $\R$ is replaced by the circle $S^1=\R/\Z$ obtained by factoring out the group of integers $\Z$. This means that a periodicity condition $\theta + 2\pi\sim\theta$ is imposed on the coordinate $\theta$ in \eqref{2.1}-\eqref{2.7}. In what follows, we will consider  precisely this group and call it the Weyl-Heisenberg group.

The multiplication law in the group $W\!H_3$ has the form
\begin{equation}\label{3.1}
(p_1, x_1, e^{\im\theta_1})\circ (p_2, x_2, e^{\im\theta_2})=
\Bigl(p_1+p_2, x_1+x_2, e^{\im\bigl(\theta_1+\theta_2+\sfrac12(p_1\cdot x_2 - p_2\cdot x_1)\bigr)}\Bigr)\ .
\end{equation}
The Lie algebra of this group has the same generators as the group $H_3$. Formulae \eqref{2.4}-\eqref{2.7} remain unchanged; the only difference is that $\theta$ in them is now an angular variable on the circle $S^1$.

\noindent{\bf Principal bundle $P(\R^6, \Uv)$.} The center of group $W\!H_3$ is group $\Uv$, and $W\!H_3/\Uv =\R^6$ holds. By defining the obvious projection
\begin{equation}\label{3.2}
\pi : \quad W\!H_3\ \stackrel{\Uv}{\longrightarrow}\ \R^6\ ,
\end{equation}
we see as a manifold the group $W\!H_3$ is a principal $\Uv$-bundle over the symplectic space
\begin{equation}\label{3.3}
W\!H_3=P(\R^6, \Uv)\ ,
\end{equation}
with projection \eqref{3.2} onto the base $\R^6$ and structure group $\Uv$.

\noindent{\bf Connection and curvature.} The connection on the bundle \eqref{3.3} is not arbitrary. It is fixed by the geometry of the group $W\!H_3$, the Maurer-Cartan equations \eqref{2.3}-\eqref{2.5}, and the Heisenberg group cocycle $\omega$,
\begin{equation}\label{3.4}
\omega =\dd A^\theta\with A^\theta =\sfrac12(p_a\dd x^a - x^a\dd p_a)\ .
\end{equation}
Note that $A^\theta$ is defined up to a gauge transformation; $A^\theta$ can be replaced by $A^\theta+\dd\vph$, where $\vph$ is a function of $x$ and $p$.

The covariant derivatives in the bundle \eqref{3.2} are given by the vector field \eqref{2.6},
\begin{equation}\label{3.5}
\nabla_{x^a}^{}=\dpar_{x^a}^{}+A_{x^a}^{\theta}\Jv\und \nabla_{p_a}^{}=\dpar_{p_a}^{}+A_{p_a}^{\theta}\Jv
\end{equation}
The curvature of the connection \eqref{3.4} is
\begin{equation}\label{3.6}
\Fv=F^\theta\Jv =\omega\Jv\for\Av=A^\theta\Jv\ ,
\end{equation}
where $\Jv =-\dpar_\theta$ is a generator of the group $\Uv$. 

\section{Line bundles $L^{(r)}$ associated with $W\!H_3$}

\noindent{\bf Characters of group $\Uv$.}     Note that the structure group  $\Uv$ of the bundle \eqref{3.2}-\eqref{3.3} is a subgroup of the 
 Weyl-Heisenberg group $W\!H_3$ and has no relation to the group U(1)$_{\sf{em}}^{}$ of electromagnetic interactions. That is why the background gauge field $\Av$ from \eqref{3.4}-\eqref{3.6} determines specific quantum interactions that lie at the foundation of quantum physics. Possessing a charge responsible for interaction with field $\Av$ is precisely what it means to be a ``quantum"  field.

All irreducible complex representations of the group $\Uv$ are one-dimensional; in other words, they are defined on the space $\C$. Each irreducible representation $\rho_r$: $\Uv\to\sGL(1, \C)$ maps an element $e^{\im\theta}$ of group $\Uv$ to an element of group GL(1,$\C$) according to the rule
\begin{equation}\label{4.1}
\rho_r(e^{\im\theta})=e^{\im r\theta}\ ,\quad r\in\Z\ .
\end{equation}
The trace of the representation matrix is called the character of the representation, and for group $\Uv$ the character is $\chi_r=\exp(\im r \theta)$, where $r$ is an integer.

\noindent{\bf Associated vector bundles $L^{(r)}$.} We have shown that the group $W\!H_3$ is a principal fibre bundle \eqref{3.3} with structure group $\Uv$. To each character $\chi_r^{}$ of this group one can associate a complex line bundle 
\begin{equation}\label{4.2}
L^{(r)} = W\!H_3\times_{\chi_r}^{}\C
\end{equation}
using the equivalence relation
\begin{equation}\label{4.3}
WH_3{\times}\C\ni (g,y)\sim (ge^{\im\theta}, e^{-\im r\theta}y)\in W\!H_3{\times}\C\ ,
\end{equation}
where $g\in W\!H_3$ and $\chi_r\in\Uv$. Here $y$ is a complex coordinate on fibre of the bundle
\begin{equation}\label{4.4}
\pi_r:\quad L^{(r)}\stackrel{\C}{\longrightarrow}\R^6\ ,
\end{equation}
and $\pi_r$ is the obvious projection onto the base $\R^6$. Thus, a countable number of complex line bundles $L^{(r)}$ is associated with the group $W\!H_3$.

\noindent{\bf Sections of the bundle $L^{(r)}$.} The sections of the bundle \eqref{4.4} are functions $f^{(r)}$ on the group $W\!H_3$ satisfying the condition
\begin{equation}\label{4.5}
f^{(r)}(ge^{\im\theta})=e^{-\im r\theta}f^{(r)}(g)\quad\Rightarrow\quad
f^{(r)}(x,p,\theta)=e^{-\im r\theta}\psi^{(r)}(x,p)\ .
\end{equation}
These functions transform correctly under shifts $\theta\mapsto\theta +\al$, $\al\in\R$.

On sections \eqref{4.5} of the bundle \eqref{4.4}, vector fields \eqref{2.6} act as follows:
\begin{equation}\label{4.6}
\begin{split}
\nabla_{x^a}^{}(e^{-\im r\theta}\psi^{(r)})&=e^{-\im r\theta}(\dpar_{x^a}^{} +\im r A_{x^a}^{\theta})\psi^{(r)}=:e^{-\im r\theta}\nabla_{x^a}^{(r)}\psi^{(r)}\ ,
\\[2pt]
\nabla_{p_a}^{}(e^{-\im r\theta}\psi^{(r)})&=e^{-\im r\theta}(\dpar_{p_a}^{} +\im r A_{p_a}^{\theta})\psi^{(r)}=:e^{-\im r\theta}\nabla_{p_a}^{(r)}\psi^{(r)}\ ,
\\[2pt]
\nabla_{\theta}^{}(e^{-\im r\theta}\psi^{(r)})&=\Jv (e^{-\im r\theta}\psi^{(r)})=e^{-\im r\theta}(\im r\psi^{(r)})=:e^{-\im r\theta}\im\Qv\psi^{(r)}\ .
\end{split}
\end{equation}
Here, we have introduced the charge operator
\begin{equation}\label{4.7}
\Qv=-\im\Jv : \quad \Qv\psi^{(r)}=r\psi^{(r)}\ ,
\end{equation}
and the index ``$r$" on the complex ``matter" field $\psi^{(r)}$ denotes the charge of this field characterizing its interaction with the vacuum field $\Av$.

\noindent{\bf Covariant derivatives.} Usually, covariant derivatives are defined as operators $\nabla^{(r)}$ from \eqref{4.6},  rather than as vector fields \eqref{3.5}. They take the form
\begin{equation}\label{4.8}
\nabla_{x^a}^{(r)}=\dpar_{x^a}^{}+\im r A_{x^a}^{\theta}\und \nabla_{p_a}^{}=\dpar_{p_a}^{}+\im r A_{p_a}^{\theta}\ ,
\end{equation}
where the components of the background gauge field  $\Av$ on $\R^6$ are written out in \eqref{3.6}. We call field $\Av$ a vacuum field because it has no sources and coincides with potential $A^\theta$, which defines cocycle $\omega$ in  \eqref{3.4}.

\noindent{\bf Line bundles $L_\C^\pm$.} On  sections $\psi^{(1)}$ of the complex line bundle $L^{(1)}$ from \eqref{4.4}, we define a Hermitian metric
\begin{equation}\label{4.9}
\bigl(\psi^{(1)},  \psi^{(1)} \bigr)=\bar\psi^{(1)}\psi^{(1)}\ ,
\end{equation}
where the overbar denotes complex conjugation. The bundle $L^{(1)}$, equipped with this metric, becomes a Hermitian bundle, which we denote by $L_\C^+$. The metric \eqref{4.9} allows one to canonically identify the bundle $L^{(-1)}$ with the bundle $L_\C^-:=\bar L_\C^+$ that is complex conjugate to $L_\C^+$. These are two basis bundles associated with the group $W\!H_3$ since all other bundles can be described in terms of tensor product of these basis ones,
\begin{equation}\label{4.10}
L^{(n)}{=}\bigl(L_\C^+\bigr)^{\otimes n}{=}L_\C^+\otimes...\otimes L_\C^+\und
L^{(-n)}\cong\bar L^{(n)}{=}\bigl(L_\C^-\bigr)^{\otimes n}{=}L_\C^-\otimes...\otimes L_\C^-\ ,
\end{equation}
where $n\in\Nbb =(1,2,...)$.

Sections of bundles $L_\C^\pm$ will be denoted by $\psi_\pm$. The covariant derivatives \eqref{4.6} when acting on $\psi_\pm$ have the form
\begin{equation}\label{4.11}
\nabla_{x^a}^\pm =\dpar_{x^a}^{}+\im q_\pm^{}A_{x^a}^\theta\und
\nabla_{p_a}^\pm =\dpar_{p_a}^{}+\im q_\pm^{}A_{p_a}^\theta\ ,
\end{equation}
where the charges $q_\pm =\pm 1$ of the fields $\psi_+$ and $\psi_-$ are opposite. Accordingly, the curvature tensors of these bundles have the form $F^\pm_{\sf vac} =\im q_\pm\omega$, and their signs are opposite. The charges of sections $\psi_\pm^{(n)}$ of the bundles \eqref{4.10} are respectively equal to $q_\pm^{(n)}=nq_\pm^{}$.

\section{Representations of the group $W\!H_3$}

\noindent{\bf Complex coordinates.}  Note that the group $W\!H_3$ has two basic coadjoint orbits: $\R_+^6=(\R^6, \omega_+=\omega)$ and $\R_-^6=(\R^6, \omega_-=-\omega)$. These orbits are bases of the bundles $L_\C^+$ and $L_\C^-$. In fact, the charges $q_\pm^{}=\pm 1$ are the signs in front of the symplectic form $\omega$ in $\R_\pm^6$. On these two spaces we introduce complex coordinates
\begin{equation}\label{5.1}
z_\pm^a=\frac{1}{\sqrt 2}(x^a\mp\im p^a)\ ,\quad 
\dpar_{z_\pm^a}^{}=\frac{\dpar}{\dpar z_\pm^a}=\frac{1}{\sqrt 2}(\dpar_{x^a}\pm\im \dpar_{p^a})\ ,
\end{equation}
where the indices $a,b=1,2,3$ are raised and lowered using Kronecker symbols. Note that $z_-^a$ are complex conjugate of $z_+^a$. However,  we use the notation \eqref{5.1} because it is convenient for unifying the consideration of the bundles $L_\C^+$ and $L_\C^-$ conjugate to each other.

\noindent{\bf Covariant derivatives.} Using formulae \eqref{4.11}, we introduce covariant derivatives of type (1,0) and (0,1) in the bundles $L_\C^\pm$. Namely, we introduce operators
\begin{equation}\label{5.2}
\nabla_{z_\pm^a}=\frac{1}{\sqrt 2}\bigl(\nabla_{x^a}^{\pm}\pm\im\nabla_{p_a}^{\pm}\bigr)=\dpar_{z_\pm^a}+\frac{\bar z_\pm^a}{2}\ ,
\quad
\nabla_{\bar z_\pm^a}=\frac{1}{\sqrt 2}\bigl(\nabla_{x^a}^{\pm}\mp\im\nabla_{p_a}^{\pm}\bigr)=\dpar_{\bar z_\pm^a}-\frac{z_\pm^a}{2}\ ,
\end{equation}
acting on smooth sections $\Psi_\pm$ of bundles $L_\C^\pm$. 

\noindent{\bf Holomorphicity/antiholomorphisity.} We introduce Dolbeault operators
\begin{equation}\label{5.3}
\bar\dpar_{L_\C^\pm}^{}=\dd\bar z_\pm^a\bigl(\dpar_{\bar z_\pm^a}+\frac{z_\pm^a}{2}  \bigr)\ ,
\end{equation}
and impose conditions
\begin{equation}\label{5.4}
\bar\dpar_{L_\C^\pm}^{}\Psi_\pm =0
\end{equation}
on the sections $\Psi_\pm$. These equations define conditions for the holomorphy of sections $\Psi_\pm$ in the trivialization under consideration. 

The resolution of conditions \eqref{5.4} are functions
\begin{equation}\label{5.5}
\Psi_\pm =\psi_\pm (z_\pm)\psi_0\ ,
\end{equation}
where $\psi_\pm (z_\pm)$ are holomorphic functions of $z_\pm^a$. The function
\begin{equation}\label{5.6}
\psi_0=\exp(-\sfrac12|z|^2)=\exp(-\sfrac12\delta_{ab} z^a\bar z^b)   
\end{equation}
defines a Hermitian metric on the bundle $L_\C^\pm$,
\begin{equation}\label{5.7}
(\Psi_\pm , \Psi_\pm)=\Psi_\pm^*\Psi_\pm=\psi_\pm^*\psi_\pm^{}e^{-|z|^2}\ ,
\end{equation}
where ``$\ast$" (together with overbar) defines complex conjugation. Recall that $z_-=\bar z_+=\bar z$, so $\psi_-(z_-)=\psi_-(\bar z)$ is an antiholomorphic function of $z$.

\noindent{\bf Hilbert spaces $\CH_\pm$.} Using \eqref{5.7}, we introduce the inner products on spaces $\CH_\pm$ of holomorphic sections of bundles $L_\C^\pm$ as integrals
\begin{equation}\label{5.8}
\langle \Psi_\pm , \Psi_\pm^\prime  \rangle=\int_{\C^3}\psi_\pm^*(z_\pm)\psi_\pm^\prime(z_\pm)e^{-|z|^2}\dd^3z <\infty\ .
\end{equation}
We get two spaces $\CH_\pm$ of holomorphic square-integrable functions. On these Hilbert spaces $\CH_+$ and $\CH_-$, two unitarily nonequivalent representation $\pi_1$ and $\pi_{-1}$ of the group $W\!H_3$ are defined.

\noindent{\bf Creation and annihilation operators.} It is easy to verify that the covariant derivative \eqref{5.2} act on sections \eqref{5.5} of the bundles $L^\pm_\C$ as follows:
\begin{equation}\label{5.9}
\begin{split}
\nabla_{z^\pm_a}^{}(\psi_\pm\psi_0)=\bigl(\dpar_{z^\pm_a}\psi_\pm)\psi_0,\quad \nabla_{\bar z_\pm^b}(\psi_\pm\psi_0)=-(z^\pm_b\psi_\pm)\psi_0
\\[2pt]
\Rightarrow\quad c_\pm^a:=\nabla_{z^\pm_a}^{},\quad       c_{b\pm}^\dagger:=-\nabla_{\bar z_\pm^b}^{},\quad 
\bigl [c_\pm^a, c_{b\pm}^\dagger\bigr ]=\delta_b^a\ .
\end{split}
\end{equation}
From these formulae we see that the covariant derivatives $\nabla^{}_{\bar z_\pm^a}$ and $\nabla^{}_{z_\pm^a}$ with connection $\Av\in\mbox{Lie}\Uv$ are creation and annihilation operators in Hilbert spaces $\CH_\pm$.

\noindent{\bf Summary.} Up to this point, we described the following chain:
\begin{equation}\label{5.10}
H_3\to W\!H_3\to P(\R^6, U(1)_{\sf v})\to L_\C^\pm\to\CH_\pm .
\end{equation}
We have shown that the Weyl-Heisenberg group $W\!H_3$ defines two gauge theories with background gauge field $\Av$ on the associated complex line bundles $L_\C^\pm$. The field $\Av$ is not arbitrary, its field strength coincides with the symplectic two-form $\omega$, which is a cocycle of the Heisenberg group. Holomorphic sections $\Psi_\pm$ of bundles $L_\C^\pm$ have charges $q_\pm =\pm 1$ that define their interaction with the field $\Av$ through covariant derivatives. The covariant derivatives \eqref{5.2} themselves, due to interaction of $\Psi_\pm$ with $\Av$, become creation and annihilation operators, generating Hilbert spaces $\CH_\pm$ from the ground state vector \eqref{5.6}. These spaces $\CH_\pm$ carry inequivalent unitary representations of the Heisenberg group.

\section{Quantum particles and antiparticles}

\noindent{\bf Particles$\,\&\,$antiparticles.} Let us emphasize that we have not ``quantized" anything, since all necessary ``quantum" objects are present in the geometry of the Weyl-Heisenberg group and the associated complex line bundles $L_\C^\pm$. From quantum mechanics it is known that $\R_+^6=(\R^6, \omega )$ is the space of states of a classical particle, and the Hilbert space $\CH_+$ is the space of states of the corresponding quantum particle. We have shown that the wave function $\psi_+\in\CH_+$ has a quantum charge $q_+=1$ associated with the character $\chi_1$ of the (representation of) central subgroup $\Uv$ of group $W\!H_3$. This holomorphic (in the Bargmann-Segal representation) function $\psi_+$ describes the particles. It is natural to ask what describes the space $\CH_-$ of antiholomorphic functions $\psi_-\in\CH_-$ with quantum charge $q_-=-1$. The answer is obvious -- these are quantum antiparticles corresponding to the classical space of states $\R_-^6=(\R^6, -\omega)$. The mapping $\omega\mapsto -\omega$ is equivalent to the time mapping $t\mapsto -t$ and is given by the charge conjugation operator $\hat C$. This is an antilinear antiunitary operator $\hat C$ mapping the spaces $\CH_+$ and $\CH_-$ into each other. It is well known both in the theory of representations of Lie groups and in quantum mechanics.

\noindent{\bf Condensate of particles.} Differential geometry provides an unambiguous answer not only to the question of interpreting the sections of the bundle $L_\C^-$, but also to the question of understanding the sections of the tensor products of bundles \eqref{4.10}. If the bundle $L_\C^+$ describes one particle, then its tensor power $(L_\C^+)^{\otimes n}$ describes $n$ identical particles, each of which is in the same quantum state. The function $|\psi_+^{(n)}|^2$ describes the particle number density normalized to $n$. The function $|\psi_+^{(n)}|^2$ can, for example, be understood as the integral of the function $|\psi_+(x_1,...,x_n)|^2$, defined on the space $\R^{3n}$, over the coordinates of all particles except one. This condensate can also be reached through the Gross-Pitaevskii equation.

\noindent{\bf Classical limit.} Until now, throughout the paper we have used formulae with dimensionless coordinates $x,p$ and $\theta$ as is customary in mathematics. To move to dimensional coordinates, one should enter the characteristic length $x_0$, the characteristic momentum $p_0$ and introduce the fundamental Planck constant using the formula $\hbar =p_0x_0$. Then the covariant derivative \eqref{4.11} will have the same form, but the charges $q_\pm$ become dimensional constant:
\begin{equation}\label{6.1}
q_\pm =\pm\frac{1}{\hbar}\ .
\end{equation}
In this case, the gauge transformation of the connection $\Av$ will be given by exponents of the form $\exp(\im S/\hbar)$, where the function $S$ has the dimension of action.

In dimensional coordinates, sections $\psi_+^{(n)}$ of the complex line bundle $(L_\C^+)^{\otimes n}$, describing the condensate of $n$ particles, have charge
\begin{equation}\label{6.2}
q_+^{(n)} =nq_+=\frac{1}{\hbar_{(n)}^{}}\with \hbar_{(n)}^{}= \frac{\hbar}{n}   \ .
\end{equation}
The transition from the wave function $\psi_+^{(1)}$ to $\psi_+^{(n)}$ is equivalent to replacing the constant $\hbar$ with the ``effective" Planck constant $\hbar_{(n)}$ in all formulae. Planck's fundamental constant $\hbar$ is fixed and cannot decrease. The transition from quantum to classical mechanics corresponds to the transition from a single-particle $\psi_+^{(1)}$ to an $n$-particle $\psi_+^{(n)}$ model and an increase in $n$ to $n>>10^{20}$. Adding interaction with the environment (decoherence) to the condensate of $n$ particles with large $n$ completes the transition to classical physics. There is no boundaries between the quantum and classical realms, the difference is only in the number of particles.

\noindent{\bf Operators $\hat p^\pm$, $\hat x_\pm$ and CCR.} We have described the representation $\pi_\hbar$ and $\pi_{-\hbar}$ of the Heisenberg group on the spaces $\CH_+$ and $\CH_-$ of holomorphic functions $\psi_+$ (particles) and antiholomorphic functions $\psi_-$ (antiparticles). By virtue of the Stone-von Neumann theorem, these representations are unitarily equivalent to representations on spaces $L_\pm^2(\R^3)$ of square-integrable functions of coordinates $x^a$ or of momenta $p_a$. Moreover, there exist isometric bijective operators $W_\pm: \CH_\pm\to L_\pm^2(\R^3)$ that define this equivalence (the Segal-Bargmann transform \cite{Segal, Barg, Hall}). In fact, the spaces $L_+^2(\R^3)$ and $L_-^2(\R^3)$ are the same as function space, but the canonical commutation relations (CCR) have a complex conjugate form on them.

The following operators are defined on spaces $L_\pm^2(\R^3)$:
\begin{equation}\label{6.3}
\hat p_a^\pm=-\frac{\im}{q_\pm}\nabla^\pm_{x^a}\ ,\quad  \hat x^a_\pm=-\frac{\im}{q_\pm}\nabla^\pm_{p_a}\ \Rightarrow\ 
\bigl [\hat x_\pm^a, \hat p_b^\pm\bigr ]=\frac{\im}{q_\pm}\delta_b^a\ .
\end{equation}
In dimensional variables we have
\begin{equation}\label{6.4}
\begin{split}
\hat p_a^+=-\im\hbar\nabla_{x^a}^+\ ,\quad \hat x^a_+=-\im\hbar\nabla_{p_a}^+\ ,\quad 
\bigl [\hat x_+^a, \hat p_b^+\bigr ]=\im\hbar\delta_b^a\ ,
\\[2pt]
\hat p_a^-=\im\hbar\nabla_{x^a}^-\ ,\quad \hat x^a_-=\im\hbar\nabla_{p_a}^-\ ,\quad 
\bigl [\hat x_-^a, \hat p_b^-\bigr ]=-\im\hbar\delta_b^a\ .
\end{split}
\end{equation}
The choice of representation is related with the choice of one or another gauge for the background gauge field $\Av$. For example, for the position representation, $\Av=-\im x^a\dd p_a$ is chosen and for the momentum representation, $\Av=\im p_a\dd x^a$. As a result, for three different representations we have
\begin{equation}\label{6.5}
\begin{array}{lll}
\mbox{complex}&\quad\mbox{position}&\quad\mbox{momentum}
\\[3pt]
\hat z_+^a=z^a=\frac{1}{\sqrt 2}\left (x^a-\frac{\im x_0}{p_0}p^a\right)&\quad
\hat x_+^a=x^a,\ \hat p_a^+=-\im\hbar\dpar_{x^a}&\quad
\hat x^a_+=\im\hbar\dpar_{p_a} ,\ \hat p_a^+=p_a
\\[3pt]
\hat z_-^a=\bar z^a=\frac{1}{\sqrt 2}\left (x^a+\frac{\im x_0}{p_0}p^a\right)&\quad
\hat x_-^a=x^a ,\ \hat p_a^-=\im\hbar\dpar_{x^a}&\quad
\hat x^a_+=-\im\hbar\dpar_{p_a} ,\ \hat p_a^-=p_a
\\[3pt]
\bigl [\dpar_{z_\pm^a}^{}, z^b_\pm\bigr ]=\delta^b_a&\quad
\bigl [\hat x_\pm^a, \hat p_b^\pm\bigr ]=\pm\im\hbar\delta_b^a&\quad
\bigl [\hat x_\pm^a, \hat p_b^\pm\bigr ]=\pm\im\hbar\delta_b^a
\end{array}
\end{equation}
Here $x_0$ and $p_0$ are the characteristic length and characteristic momentum (e.g. for the harmonic oscillator $x_0=\sqrt{\hbar/m\omega}$ and $p_0=\sqrt{\hbar m\omega}$).

\noindent{\bf Energy operators $\hat E_\pm$.} It is easy to see that in the position representations the eigenvectors of the momentum operators are the following functions:
\begin{equation}\label{6.6}
\hat p_a^\pm e^{\im q_\pm p\cdot x}=p_ae^{\im q_\pm p\cdot x}\for q_\pm=\pm\frac{1}{\hbar}\ ,\quad p\cdot x=p_ax^a\ .
\end{equation}
Similar eigenfunctions can be written for operators $\hat x_\pm^a$ in the momentum representation. It is formulae \eqref{6.3}-\eqref{6.6} that distinguish between spaces $L_+^2(\R^3)$ and $L_-^2(\R^3)$.

Note that the energy operators $\hat E_+$ and $\hat E_-$ on the spaces $L_+^2(\R^3)\cong\CH_+$ and $L_-^2(\R^3)\cong\CH_-$, like the operators $\hat p_a^+$ and $\hat p_a^-$, are complex conjugate to each other:
\begin{equation}\label{6.7}
\hat E_+=\im\hbar\dpar_t\und\hat E_-=-\im\hbar\dpar_t\ .
\end{equation}
They are generators of translation in time in spaces $L_\pm^2(\R^3)$. From formulae \eqref{6.7} follows the definition of frequency
\begin{equation}\label{6.8}
\hat E_\pm e^{-\im q_\pm Et}= Ee^{-\im q_\pm Et}\ ,
\end{equation}
and it can be seen that the energy is the same for particle and antiparticles. By combining the formulae \eqref{6.6}-\eqref{6.8} we arrive at positive- and negative-frequency exponents
\begin{equation}\label{6.9}
e^{-\im q_\pm(Et-p_ax^a)}=e^{\mp\frac\im\hbar (Et-p_ax^a)}     \ ,
\end{equation}
that distinguish particles and antiparticles in Minkowski space. From the formulae \eqref{6.6}-\eqref{6.8} it follows that the component $p_0$ of 4-momentum for free antiparticles is negative, but their energy $E=-p_0c$  is positive. There is no gap between the energies of particles and antiparticles, their energies are always equal. We also recall that the spaces $\CH_+\cong L_+^2(\R^3)$ and $\CH_-\cong L_-^2(\R^3)$, as well as the CCR, are related by the charge conjugation map
\begin{equation}\label{6.10}
\hat C:\quad \zh^a\mapsto\hat\zb^a,\quad \hat E_+\mapsto\hat E_-,\quad \ph_a^+\mapsto\ph_a^-,\quad\omega\mapsto-\omega\und t\mapsto -t\ .
\end{equation}
Particles and antiparticles are present in the theory absolutely symmetrically, and the concept of an antiparticle is in no way connected with the transition to a relativistic theory. The concept of particle and antiparticle is unambiguously related to the subgroup $\Uv$ of the Weyl-Heisenberg group $W\!H_3$.

\noindent{\bf General case.} We have considered the canonical case when the phase space $\R^6$ of a classical particle is a homogeneous space of the group $W\!H_3$ with respect to the subgroup $\Uv$. However, from the geometric quantization approach it is known \cite{Sour}-\cite{Wood}  that quantum mechanics can be defined on a wide class of symplectic manifolds $X^6$ (cotangent bundles, K\"ahler manifolds, and so on). How to define particles and antiparticles in this case? The answer is that by Darboux's theorem, locally, on sufficiently small open sets $U_\al\subset X^6$, we still have $U_\al=W\!H_3(U_\al)/\Uv$ and can introduce spaces $\CH_\pm (U_\al)$ of functions that are holomorphic and antiholomorphic on $U_\al$. Functions $\psi_\pm^{U_\al}\in\CH_\pm(U_\al)$ have opposite weights relative to the group $\Uv\subset W\!H_3(U_\al)$. After choosing a covering $\{U_\al\}$ of the manifold $X^6$, we sew together all these locally defined functions $\psi_\pm^{U_\al}$ into a globally defined sections of the bundles $L_\C^\pm$. Although the Weyl-Heisenberg group is not a symmetry group acting on the manifold $X^6$, its subgroup $\Uv$ remains the structure group of the bundle $P(X^6, \Uv)$ generalizing the bundle \eqref{3.3} and the complex line bundles $L_\C^\pm$ associated with it. This is proven in the geometric quantization approach \cite{Sour}-\cite{Wood}.

\section{Schr\"odinger equations}

\noindent{\bf Hamiltonians.}  We discussed the state spaces of classical and quantum particles and antiparticles. In both cases, symplectic two-forms $\omega_{\pm}$ on the phase space play an important role. Particles and antiparticles differ in the sign of these 2-forms $\omega_{\pm}=\pm\omega$ on the orbits $\R^6_+=(\R^6, \omega)$ and $\R^6_-=(\R^6, -\omega)$ of the coadjoint action of the Heisenberg group. This sign is also related to the sign of complex structures $\pm\CJ$ on $\R^6$ and holomorphic/antiholomorphic structures on complex line bundles $L_\C^\pm$ over $\R^6$.

The changes of classical states over time are given by one-parameter groups generated by Hamiltonian vector fields
\begin{equation}\label{7.1}
X_H^\pm =\omega_\pm^{-1}(\dd H) =\pm\left(\frac{\dpar H}{\dpar p_a} \frac{\dpar}{\dpar x^a}  - 
\frac{\dpar H}{\dpar x^a} \frac{\dpar}{\dpar p_a}\right) \ ,
\end{equation}
where $H(x, p)$ is the energy function. The change of quantum states is determined by one-parameter groups generated by Hamiltonian operators $\hat H(\xh_\pm, \ph_\pm )$ constructed using the function $H$. These groups act in spaces $L_\pm^2(\R^3)$ or $\CH_\pm$  depending on the chosen representation. We consider time-independent functions $H$ and operators $\hat H_\pm$ (energy conservation), and we choose for them the standard form 
\begin{equation}\label{7.2}
\hat H_\pm =\frac{1}{2m}\,\delta^{ab}\,\ph_a^\pm\,\ph_b^\pm + V(\xh )\ .
\end{equation}
Here $\ph_a^\pm$ are the kinetic momentum operators, that is, quantum version of the momenta $p_a=m \dd x^a/\dd t$. For the symplectic form \eqref{3.4} they coincide with the canonical momenta that, after quantization, go over to the operators \eqref{6.5} conjugate to the coordinate operators.

\noindent{\bf Electromagnetic field.} In symplectic geometry, the electromagnetic field is introduced by adding the two-form of the field strength $F_{\sf em}$ to the two form $\omega_\pm$ of the canonical symplectic form \cite{Sour, Stern, Mont}
\begin{equation}\label{7.3}
\omega_\pm =\dd P_a^\pm\wedge\dd x^a\mapsto \Omega_\pm =\dd\Pi_a^\pm\wedge\dd x^a + e F_{\sf em}\ .
\end{equation}
Note that $P_a^-=-P_a^+$ but $\Pi_a^-\ne -\Pi_a^+$. With such a change in the symplectic form, the kinetic momenta $\Pi_a^\pm$ cease to commute with respect to the Poisson bracket and, when quantized, are mapped into the covariant derivatives $D_a^\pm$ in the position representation. That is why, we relabeled them as $\Pi_a^\pm$, leaving the notation $P_a^\pm$ and $p_a^\pm$ for commuting canonical momenta.

Formula \eqref{7.3} can be rewritten as
\begin{equation}\label{7.4}
\Omega_\pm=\dd\Pi_a^\pm\wedge\dd x^a + e \dd (A_a^{\sf em}\dd x^a)=\dd\Bigl [(\Pi_a^\pm + eA_a^{\sf em})\dd x^a\Bigr]=\dd p_a^\pm\wedge \dd x^a\ ,
\end{equation}
\begin{equation}\label{7.5}
\Rightarrow\ \Pi_a^\pm =p_a^\pm -eA_a^{\sf em}\Rightarrow\ \hat\Pi_a^\pm =\mp\im\hbar\bigl(\dpar_{x^a}^{}\pm \frac{\im e}\hbar A_a^{\sf em}\bigr)=:\mp\im\hbar D_a^\pm\ .
\end{equation}
In this case, the Hamiltonians \eqref{7.2} are replaced by operators
\begin{equation}\label{7.6}
\hat  H_\pm = -\frac{\hbar^2}{2m}\,\delta^{ab}\,D_a^\pm D_b^\pm + V(x)\ ,
\end{equation}
where $D_a^\pm$ correspond to kinetic momenta, which do not commute with each other when $A_a(x){\ne}0$. These operators act on $\psi_\pm$ and from  \eqref{7.6} it is clear that $\psi_+$ and $\psi_-$ have opposite electric charges $e_+=e$ and $e_-=-e$.

\noindent{\bf Charges $(nq_\pm , ke_\pm )$}. In the language of differential geometry, electric charges $e_\pm =\pm e$ associated with field $\psi_\pm$ are introduced as follows. The projection $\pi$: $\R^6\to\R^3$ of the phase space $\R^6$ onto the coordinate space $\R^3$ is considered. Over the space $\R^3$, a principal bundle $P(\R^3, \sU(1)_{\sf em})$ is introduced and two complex conjugate line bundles 
$E_{\sf em}^\pm$ are associated with it. Using the projection $\pi$, they are pulled back from $\R^3$ to the line bundles $\pi^*E_{\sf em}^\pm$ over $\R^6$. After this, the tensor products $L_\C^+\otimes \pi^*E_{\sf em}^+$ and $L_\C^-\otimes \pi^*E_{\sf em}^-$ are introduced. Although the fibres of these  bundles are complex one-dimensional spaces, their sections $\psi_{++}$ and $\psi_{--}$ carry two indices corresponding to charges $(q_+, e_+)$ and $(q_-, e_-)$  with respect to the group $\sU(1)_{\sf v}\times\sU(1)_{\sf em}$. It is important to understand that these fields interact with varying intensities (determined by charges) with different gauge fields $\Av$ and $A_{\sf em}$. This can be clearly seen if we move from unit charges to charges $(n q_\pm , ke_\pm )$ of sections $\psi_{\pm\pm}^{(n,k)}$ of bundles $(L_\C^\pm)^{\otimes n}\otimes (\pi^*E_{\sf em}^\pm)^{\otimes k}$. In particular, when $k=0$, the electromagnetic field is switched off,  but the field $\Av$ is not, since in quantum mechanics $n{\ne}0$, and classical mechanics is obtained when $n$ tends to infinity. In the single-particle case, for the sake of notational simplicity, we retain only one index for sections $\psi_\pm$.

\noindent{\bf Schr\"odinger equation for $\psi_+$ and $\psi_-$.} The change in state vectors $\psi_\pm\in L_\pm^2(\R^3)\cong\CH_\pm$ over time is determined by different evolution operators
\begin{equation}\label{7.7}
U_\pm (t)=e^{-\im t q_\pm\hat H_\pm}\ .
\end{equation}
These operators define one-parameter subgroups in the unitary group $\CU(L_\pm^2(\R^3))$ acting on the spaces $L_\pm^2(\R^3)$. By defining the evolution of $\psi_\pm$ using the operators \eqref{7.7}, we obtain two Schr\"odinger equations,
\begin{equation}\label{7.8}
\psi_\pm (t) = U_\pm (t)\psi_\pm (0)\quad\Rightarrow\quad\pm\im\hbar\dpar_t\psi_\pm =\hat H_\pm   \psi_\pm   \ ,
\end{equation}
the first for particles and the second for antiparticles. Note that on the left side of these equations we see two energy operators $\hat E_\pm =\pm\im\hbar\dpar_t$.

Obviously, for stationary states of particles and antiparticles we obtain the same energy values,
\begin{equation}\label{7.9}
\psi_\pm = e^{-\im q_\pm E t}\tilde\psi_\pm\quad\Rightarrow\quad\hat H_\pm\tilde\psi_\pm =E\tilde\psi_\pm\ .
\end{equation}
This also follows from the application of the charge conjugation operator \eqref{6.10}. This operator $\hat C$ maps the representation spaces $\CH_+$ and $\CH_-$ of the group $W\!H_3$ into each other and changes the sign of the charges $(q_+, e_+)\mapsto (q_-, e_-)=-(q_+, e_+)$. This operator naturally transforms the equations for particles into equations for antiparticles with distinct charges $q_+=1/\hbar$ and $q_-=-1/\hbar$, even if the particles are electrically neutral.

\noindent{\bf Noether currents.} Let us consider the global action of the group $\Uv$ on sections,
\begin{equation}\label{7.10}
\psi_\pm\mapsto e^{\im\al_\pm}\psi_\pm\ .
\end{equation}
Equations \eqref{7.8} are invariant with respect to these transformations and, according to Noether's theorem, in this model the conserved charges $q_\pm$ are defined. Electric charges of the fields $\psi_\pm$, if introduced, will be conserved as a consequence of the conservation of the charges $q_\pm$.

Let us consider Hamiltonians \eqref{7.6} with electromagnetic fields \eqref{7.5} supplemented by extended energy operators
\begin{equation}\label{7.11}
\hat E_\pm \mapsto \hat\CE_\pm =  \hat E_\pm + eA_t^{\sf em}\quad\Leftrightarrow\quad \pm\im\,\hbar (\dpar_t\mp\im\,\sfrac{e}\hbar\, A_t^{\sf em})=:\pm\im\,\hbar \,D_t^\pm  \ .
\end{equation}
The Schr\"odinger equations then take the following form:
\begin{equation}\label{7.12}
\pm\im\,\hbar \,\dpar_t\psi_\pm =\Bigl(-\frac{\hbar^2}{2m}\,\delta^{ab}\,D_a^\pm\,D_b^\pm - eA_t^{\sf em} + V(x)\Bigr)\psi_\pm
\end{equation}
Here, we moved $eA_t^{\sf em}$ from \eqref{7.11} to the right-hand side of the equation.

From equation \eqref{7.12}, it is straightforward to derive four-vectors of current densities $(j_\pm^t, j_\pm^a)$ that satisfy the continuity equations
\begin{equation}\label{7.13}
\dpar_t j_\pm^t + \dpar_{x^a} j_\pm^a=0\ \Rightarrow\  j_\pm^t=\pm\psi_\pm^*\psi_\pm\ ,\  j_\pm^a =
\mp\frac{\im\hbar}{2m}\bigl (\psi_\pm^*D_a^\pm\psi_\pm - \psi_\pm (D_a^\pm\psi_\pm)^*\bigr )\ ,
\end{equation}
where $j_\pm^t$ represent the charge densities,
\begin{equation}\label{7.14}
q_\pm=\pm\frac1\hbar\int\psi_\pm^*\psi_\pm\dd^3x = \pm\frac1\hbar\ .
\end{equation}
From \eqref{7.12} and \eqref{7.14} we see that $j_\pm^t$ are exactly the densities of quantum charges, accurate to a dimensional constant. The interpretation of the functions
\begin{equation}\label{7.15}
\rho_\pm=|\psi_\pm|^2 = \psi_\pm^*\psi_\pm
\end{equation}
as probability densities is secondary to the formula \eqref{7.14}. We emphasize that the charges $q_+$ and $q_-$ are conserved independently of each other. In other words, the number of particles and the number of antiparticles in quantum mechanics are set initially and cannot change. There are no operators that could change these numbers.

\noindent{\bf Vector nature of $\psi_\pm$.} Historically, $\psi_\pm$ are called wave functions. However, these are not functions, but components of complex vectors of the type (1,0) and (0,1). The objects $\psi_+$ and $\psi_-$ under consideration are sections of complex conjugate {\it vector} bundles $L_\C^+$ and $L_\C^-$ of rank one. For such bundles and their sections, only the direct sum is defined. Also 
$\psi_+$ and $\psi_-$ belong to different Hilbert spaces $L_+^2(\R^3)\cong\CH_+$ and $L_-^2(\R^3)\cong\CH_-$ for which only the direct sum is defined.

Vectors $\psi_+$ and $\psi_-$ can be added only using bases $v_\pm^{}$ in the fibres of the bundles $L_\C^\pm$, 
\begin{equation}\label{7.16}
v_+=\frac{1}{\sqrt2}\,\begin{pmatrix}1\\-\im\end{pmatrix},\quad v_+=\frac{1}{\sqrt2}\,\begin{pmatrix}1\\\im\end{pmatrix},\quad
v_+^\+ v_-=v_-^\+ v_+=0\ .
\end{equation}
Using these bases, the sections of the bundles $L_\C^+$ and $L_\C^-$ can be added together using the formula
\begin{equation}\label{7.17}
\psi=\frac{1}{\sqrt2}\,\begin{pmatrix}\psi_++\psi_-\\-\im(\psi_+-\psi_-)\end{pmatrix}=\psi_+v_++\psi_-v_-\in L_\C^+\oplus L_\C^-\ .
\end{equation}
In principle, the two Schr\"odinger equations \eqref{7.8} can be combined into one matrix equations 
\begin{equation}\label{7.18}
\Jv\dpar_t\psi = \sfrac1\hbar\,\hat H\psi\for \hat H:=\hat H_+v_+v_+^\+ + \hat H_-v_-v_-^\+\ .
\end{equation}
Here
\begin{equation}\label{7.19}
\Jv = \im\,(v_+v_+^\+ - v_-v_-^\+)\ , \quad \Jv v_+=\im v_+\ ,\quad \Jv v_-=-\im v_-
\end{equation}
is the matrix representation of the generator $\Jv =-\dpar_\theta$ of the structure group $\Uv\subset W\!H_3$ of the bundle $ L_\C^+\oplus L_\C^-$. Accordingly, $\Qv =-\im\Jv$ is a matrix representation of the charge operator.

\section{Pauli equations}

\noindent{\bf Clifford algebra Cl$^\C(3)$.} It is generally believed that in the 3-dimensional space $\R^3$ of non-relativistic quantum mechanics one two-component spinor is sufficient. This is not entirely true. The point is that the Pauli matrices are generators of the Clifford algebra Cl$(3,0)\cong \mbox{Mat}(2,\C)$ of the space $\R^3$ only over the field of real numbers. This means that matrix $\sigma_1\sigma_2=\im\sigma_3$ is considered to be different from matrix $\sigma_3$, i.e. matrices and spinors can only be multiplied by real numbers. Therefore, it is correct to consider the complexified Clifford algebra Cl$^\C(3)\cong\,$Mat(2,$\C)\oplus\,$Mat(2,$\C$) with generators
\begin{equation}\label{8.1}
\tilde\gamma^a=\begin{pmatrix}\sigma^a&0\\0&-\sigma^a\end{pmatrix}\quad\Rightarrow\quad
\tilde\gamma^1\tilde\gamma^2=\begin{pmatrix}\im\sigma^3&0\\0&\im\sigma^3\end{pmatrix}\ne\im\tilde\gamma^3\ .
\end{equation}
This matrix acting on the columns
\begin{equation}\label{8.2}
\psi = \begin{pmatrix}\chi_+ v_+\\\chi_-v_-\end{pmatrix}\with
\chi_+=(\chi_+^\al)\in\C^2, \quad \chi_-=(\chi_{-\dal})\in\bar\C^2\ ,
\end{equation}
where the spinors $\chi_\pm$ take values in the bundles $L_\C^\pm$ similarly to formula \eqref{7.17} for $\psi_\pm$.

\noindent{\bf Two-component spinors.} Spinors $\chi_+$ and $\chi_-$ belong to representations ($\sfrac12$, 0) and (0, $\sfrac12$) of group SL(2, $\C$). They are related by the charge conjugation operator $\hat C$, which maps them into each other. It operates as follows:
\begin{equation}\label{8.3}
\hat C:\quad \C^2\ \ni\ \chi_+^\al\mapsto \bar\chi_+^\dal\mapsto\veps_{\dal\dot\beta}\bar\chi_+^{\dot\beta}=\bar\chi_{+\dal}\in\bar\C^2\ .
\end{equation}
We shall define on $\C^2$ a Hermitian metric $\delta_{\al\dal}$  that is invariant under the group U(2). In this setting, the spaces $\C^2$ and $\bar\C^2$ serve as fundamental and antifundamental representations of the group U(2). The representation $\C^2\ni\chi_+$ and $\bar\C^2\ni\chi_-$ can be related by mapping
\begin{equation}\label{8.4}
\C^2\ \ni\ \chi_+^\al\mapsto \delta^{\al\dal}\bar\chi_{+\dal}=\hat\chi^\al\in \bar\C^2\und
 \chi_{-\dal}\mapsto \delta_{\dal\al}\veps^{\al\beta}\bar\chi_{-\beta}=\hat\chi_{-\dal}\ .
\end{equation}
These mappings are smooth isomorphisms of the representations $\bf 2$ and $\bf{\bar 2}$ of the group SU(2), i.e. the defining representation of SU(2) is pseudo-real. However, the mappings \eqref{8.4} are not holomorphic isomorphisms, as seen by the central subgroup U(1)$\subset$U(2). In fact, the operator \eqref{8.3} maps spin-$\sfrac12$ particles and antiparticles into each other.

\noindent{\bf Pauli equations for $\chi_+$ and $\chi_-$.} Spinors $\chi_+$ and $\chi_-$ have quantum charges $q_+$ and $q_-$ being sections of the bundles $L_\C^+\otimes\C_+^2$ and $L_\C^-\otimes\C_-^2$, where $\C_+^2=\C^2$ and $\C_-^2=\bar\C^2$. After lifting $\chi_\pm$ to sections of the bundles $L_\C^\pm\otimes E^\pm_{\sf em}\otimes \C_\pm^2$ they have charges $(q_\pm , e_\pm)$. For two-component spinors with these charges, the Pauli equations have the form
\begin{equation}\label{8.5}
\begin{split}
\Bigl[\pm\im\hbar\, D_t^\pm + &\frac{\hbar^2}{2m}(\sigma^a D_a^\pm)^2\Bigr]\chi_\pm =0
\\[2pt]
\Rightarrow\ \pm\im\hbar\, \dpar_t\chi_\pm =\Bigl[- &\frac{\hbar^2}{2m}\bigl(\sigma^a (\dpar_a \pm\frac{\im e}{\hbar}A_a^{\sf em})\bigr)^2-eA_t^{\sf em}\Bigr]\chi_\pm\ .
\end{split}
\end{equation}
The equation for $\chi_+$ is the ordinary equation for an electron, and the equation for $\chi_-$ is the conjugate equation for a positron. Comparing \eqref{8.5} with \eqref{7.9}, we see that the energies of $\chi_+$ and $\chi_-$ fields are always equal, similar to the spinless case.

The equations \eqref{8.5} are invariant under global transformations of the group $\Uv$ and under local transformations of the group U(1)$_{\sf em}$. The density of the quantum charges of the fields $\chi_\pm$, 
\begin{equation}\label{8.6}
j_\pm^t =\pm\chi_\pm^\+\chi_\pm\quad \Rightarrow\quad q_\pm =\pm\frac1\hbar \,\int\chi_\pm^\+\chi_\pm\dd^3x=\pm\frac1\hbar\ ,
\end{equation}
is easily derived from $\Uv$-invariance, which is completely analogous to the spinless case \eqref{7.13} and \eqref{7.14}. Probability densities are equal to $\rho_\pm=\chi_\pm^\+\chi_\pm$. Charges $(q_+, e_+)$ and $(q_-, e_-)$ are conserved independently of each other.

\section{Standard Klein-Gordon theory}

\noindent{\bf Lagrangian and Noether charges.} The Klein-Gordon (KG) equation was introduced as a relativistic generalization of the Schr\"odinger equation  \eqref{7.8} for a free particle. Accordingly, it was assumed that the complex KG field $\phi$ generalizes the wave function $\psi_+$ of particle. This assumption immediately encountered contradictions, which we will describe below.

For the field $\phi\in\C$, the density of the Lagrange function was introduced, 
\begin{equation}\label{9.1}
\CL_\phi = \frac{\hbar^2}{2m}\Bigl(\eta^{\mu\nu}\dpar_\mu\phi^*\dpar_\nu\phi - \Bigl(\frac{mc}{\hbar}\Bigr)^2\phi^*\phi\Bigr)\ ,
\end{equation}
where $(\eta_{\mu\nu})=\mbox{diag}(1, -1, -1, -1)$ is Minkowski metric and $c$ is the speed of light. The factor $\hbar^2/2m$ in \eqref{9.1} is related to the fact that we use for $\phi$ the same dimension $[\phi]=L^{-3/2}$ as for the non-relativistic wave function $\psi$, since we associate $|\phi|^2$ with the probability density. From the invariance of the Lagrangian with respect to time shifts, an energy density is derived, 
\begin{equation}\label{9.2}
\CH_\phi = \frac{\hbar^2}{2m}\Bigl(\dpar_0\phi^*\dpar_0\phi + \dpar_a\phi^*\dpar_a\phi + \Bigl(\frac{mc}{\hbar}\Bigr)^2\phi^*\phi\Bigr)\ ,
\end{equation}
the integral of which over the space $\R^3$ yields a positive energy $E$ as a conserved Noether charge. From the invariance of the Lagrangian with respect to the $\Uv$-transformation $\phi\mapsto e^{-\im\al}\phi$, the density of another conserved Noether charge is derived,
\begin{equation}\label{9.3}
j_\phi^t =\frac{\im\hbar^2}{2mc^2}(\phi^*\dpar_t\phi - \phi\dpar_t\phi^*)\ ,
\end{equation}
which, by analogy with $j_+^t=\psi_+^*\psi_+$ for the Schr\"odinger equation, is identified with the probability density of detecting a particle at a point $x^a\in\R^3$.

\noindent{\bf Solutions $\phi_+$ and $\phi_-$.} The equation of motion following from the Lagrangian \eqref{9.1} has the form
\begin{equation}\label{9.4}
\Bigl(\eta^{\mu\nu}\dpar_\mu\dpar_\nu + \Bigl(\frac{mc}{\hbar}\Bigr)^2\Bigr)\phi=0\ .
\end{equation}
This equation is linear and its solution $\phi$ is the sum of two wave packets,
\begin{equation}\label{9.5}
\phi =\phi_++\phi_-\ ,\quad \phi_\pm (x)=\int_{H_\pm^3} e^{-\frac\im\hbar p_\mu^\pm x^\mu}\phi_\pm (p_\pm)\dd \mbox{Vol}_{H_\pm^3}^{}\ ,
\end{equation}
where $H_+^3$ and $H_-^3$ are two sheets of the hyperboloid in the momentum space,
\begin{equation}\label{9.6}
H^3=\bigl\{\eta^{\mu\nu}p_\mu p_\nu =m^2c^2\bigr\}=H_+^3(p_0^+>0)\cup H_-^3(p_0^-<0)\ .
\end{equation}
Here $\dd \mbox{Vol}_{H_\pm^3}^{}$ are Lorentz-invariant measures on these hyperboloids  $H_\pm^3\cong\sSO^+(1,3)/\sSO(3)$. Note that  $\phi_\pm(p_\pm)$ belong to the Hilbert spaces $L_\pm^2(H_\pm^3)$ of functions on hyperboloids $H_\pm^3$ with inner products given by the $\dd \mbox{Vol}_{H_\pm^3}^{}$-measure.

\noindent{\bf Problems of interpretation.} It is easy to verify that $\phi_+$ has positive energy both from the point of view of the formula \eqref{9.2} and from the point of view of the energy operator $\hat E_+=\im\hbar\dpar_t$. Also $\phi_+$ has a positive charge density $j^t_{\phi_+}$ which can be related to the probability density. Moreover, for the ansatz
\begin{equation}\label{9.7}
\phi_+=e^{-\im m c^2t/\hbar}\psi_+(t,x)\ ,
\end{equation}
the equation \eqref{9.4} in the limit $c\to\infty$ goes over into the Schr\"odinger equation for the particle $\psi_+$.

The energy of $\phi_-$ is positive both according to the formula \eqref{9.2} and from the point of view of the energy operator $\hat E_-=-\im\hbar\dpar_t$. The charge density $j_{\phi_-}^t$ from \eqref{9.3} is negative, as it should be. When passing to the non-relativistic limit, similarly to that considered in \eqref{9.7}, we obtain
\begin{equation}\label{9.8}
\phi_-=e^{\im m c^2t/\hbar}\psi_-(t,x)\quad\Rightarrow\quad -\im\hbar\dpar_t\psi_- = - \frac{\hbar^2}{2m}\delta^{ab}\dpar_a\dpar_b\psi_-\ .
\end{equation}
This is the Schr\"odinger equation \eqref{7.8} for a free antiparticle having a negative density \eqref{7.13} of the quantum charge $q_-$ written out in \eqref{7.14}. The charge density $j^t_{\phi_-}$  reduces precisely to it, as it should.

As we can see, the problems are not related to the Klein-Gordon equation itself, but to the interpretation of the field $\phi_-$ as a wave function of a particle (in fact, this is an antiparticle) and the absence of the concept of an antiparticle in non-relativistic quantum mechanics. Furthermore, as we showed in Section 7, only the direct sum is defined for the wave functions of particles and antiparticles (they are vectors). This is equivalent to the statement  that the Klein-Gordon equations must be defined for the fields $\phi_+$ and $\phi_-$ separately, rather than for their sum.

The main idea of this paper is that the concept of antiparticles is not related to the transition to a relativistic theory. This concept is related to the weights of the representations of the central subgroup $\Uv$ of the Weyl-Heisenberg group $W\!H_3$. For particles and antiparticles, these weights are equal to $\pm 1$, and their wave functions $\psi_\pm$ are sections of complex conjugate line bundles $L_\C^\pm$ over 6-dimensional phase manifolds. Even when passing from $\R^6=W\!H_3/\Uv$ to curved symplectic manifolds $X^6$, where the group $W\!H_3$ acts only locally, its subgroup $\Uv$ remains the structure group of the bundles $L_\C^\pm$ over $X^6$ \cite{Sour}-\cite{Wood}. Therefore, the weights (equivalently, quantum charges) of the wave functions are always fixed prior to the introduction of Lagrangians and equations.

\noindent{\bf Resolution of discrepancies.} The reason for the inconsistency of model  \eqref{9.1}-\eqref{9.5} is that ``it does not know" that
\begin{itemize}
\item $\phi_+$ is a section of a line bundle $L_\C^+$ over $T^*H_+^3$ with structure group $\Uv$\\
$\Rightarrow$ the group $\Uv$ acts on $\phi_+$ according to the rule $\phi_+\mapsto\exp (-\im\theta)\phi_+$\\
$\Rightarrow\phi_+$ has quantum charge $q_+=\sfrac1\hbar$, 
\item $\phi_-$ is a section of a line bundle $L_\C^-$ over $T^*H_-^3$ with structure group $\Uv$\\
$\Rightarrow$ the group $\Uv$ acts on $\phi_-$ according to the rule $\phi_-\mapsto\exp (\im\theta)\phi_-$\\
$\Rightarrow\phi_-$ has quantum charge $q_-=-\sfrac1\hbar$\ .
\end{itemize}
Due to the opposite transformation properties of the function $\phi_+$ and $\phi_-$, the Lagrangian \eqref{9.1} and the Hamiltonian \eqref{9.2} are not scalars with respect to the action  of the group $\Uv$ on the fields $\phi_\pm$. For example, the term $\phi^*\phi$ transforms as follows: 
\begin{equation}\label{9.9}
\phi^*\phi = (\phi_+^* + \phi_-^*)(\phi_+ + \phi_-)
\quad\mapsto\quad
\phi_+^*\phi_+ +e^{2\im\theta}\phi_+^*\phi_-+e^{-2\im\theta}\phi_-^*\phi_+ +\phi_-^*\phi_-\ .
\end{equation}
Non-invariant terms in \eqref{9.9} are responsible for the Klein paradox. This paradox does not exist in $\Uv$-invariant relativistic quantum mechanics, where charges $q_+$ and $q_-$ with densities $j^t_{\phi_+}$ and $j^t_{\phi_-}$ are conserved separately. It should be noted that when generalizing quantum mechanics to quantum field theory, the group $\Uv$ ceases to be a symmetry group of the theory, and one can introduce cross-terms responsible for pair creation and annihilation.

\section{Klein-Gordon particle and antiparticle}

\noindent{\bf Standard scheme of QM.} Recall that non-relativistic quantum mechanics is formulated via a chain of successive steps of the form
\begin{equation}\label{10.1}
\begin{array}{lll}\mbox{classical}\\\mbox{state space}\\(X^6,\omega_+)\end{array}\quad\Rightarrow\quad
\begin{array}{lll}\mbox{quantum}\\\mbox{state space}\\\quad\CH_+\end{array}\quad\Rightarrow\quad
\begin{array}{lll}\mbox{dynamics}\\\mbox{of wave}\\\mbox{functions}\ \psi_+\in\CH_+\end{array}\ .
\end{equation}
We have shown that the space of classical states $(X^6, \omega_-)$ with $\omega_-=-\omega_+$ corresponds to the Hilbert space $\CH_-$ of antiparticle wave functions $\psi_-$, which doubles the aforementioned chain for particles. Note that the manifolds $(X^6, \omega_\pm)$ define the initial data for the dynamics of the classical and quantum particles and these spaces do not depend in any way on the Hamiltonians defining the external forces acting on the particles.

Relativistic quantum mechanics is introduced according a completely different scheme. Free relativistic equations are specified, external forces are introduced through the extension of partial derivatives by gauge fields and Levi-Civita connections, and attempts are made to introduce Hilbert spaces $\CH_\pm$, which can only be done for free fields and in special cases such as the hydrogen atom. As a result, there is lack of understanding of what the classical space of state is, and what a relativistic particle and antiparticle are in the field of external forces. In fact,  the standard approach suggests considering  only asymptotically free particles in the infinite past and in the infinite future.

The classical phase space of particles was introduced by Souriau as a space of trajectories \cite{Sour} and was considered by Sniatycki \cite{Sni} and Woodhouse\cite{Wood}.
It is defined as follows. A function $\mu_m$, called the mass function, is introduced on the phase space $T^*\R^{1,3}\cong\R^{1,3}\times\R^{1,3}\ni (x^\mu , p_\nu)$. This function defines a 7-dimensional level surface $\mu_m(x,p)=mc^2$ and a one-parameter group of symplectomorphisms generated by the Hamiltonian vector field $Y_{\mu_m}^{}$, acting on this surface $\mu_m^{-1}(m)\subset T^*\R^{1,3}$. Quotienting by this group defines the manifold $X_{\mu_m}^6$ as an orbit space, while the group itself defines the dynamics as motion along the orbits. In the relativistic case, the classical state space $X_{\mu_m}^6$ depends on external forces specified by the function $\mu_m$ (see e.g. \cite{Sour, Sni, Wood}), which constitute a significant difference from the non-relativistic case. We will not discuss the general case here, but rather illustrate the foregoing with the example of a free spinless particle. In this case $\mu_m:=p^2/m$ is the Casimir function of the Poincare group. It defines a symplectic manifold $X_{\mu_m}^6=T^*H_+^3\cup T^*H_-^3$, which is a Kirillov-Kostant-Souriau orbit in the coadjoint action of the Poincare group \cite{Kir}, \cite{Sour}-\cite{Wood}, \cite{Popov3}. Taking the two orbits $T^*H_\pm^3$ as spaces of classical states, we introduce the spaces of quantum states $\CH_\pm$ and describe the transition to the dynamics of relativistic wave functions $\psi_\pm\in\CH_\pm$. In other words, for relativistic particles and antiparticles, we will reproduce the logical sequence \eqref{10.1} used to construct the theory in non-relativistic quantum mechanics.

\noindent{\bf Phase space $T^*H^3$.} Let us consider a relativistic phase space $T^*\R^{1,3}$ with coordinates $x^\mu$ and momenta $p_\mu$, $\mu , \nu =0,...,3$. On this space we define symplectic two-form
\begin{equation}\label{10.2}
\Omega = \frac1\hbar\,\dd p_\mu\wedge\dd x^\mu\ .
\end{equation}
A point of mass $m$ moves in this space along straight lines,
\begin{equation}\label{10.3}
x^\mu =x^\mu_0 + \frac{p^\mu}{m}\tau\ ,\quad p^\mu =m\frac{\dd x^\mu}{\dd\tau}=\eta^{\mu\nu}p_\nu =p_0^\mu\ ,
\end{equation}
where $\tau$ is proper time. Note that the proper time $\tau$ is conjugate to the rest energy of the particle $E_0=mc^2$, and this is significant difference from the non-relativistic case where the time $t$ is conjugate to the energy $E$. To simplify the formulae, we will consider dimensionless coordinates and parameters here, $\hbar =c=1$. The momenta belong to the two-sheeted hyperboloid \eqref{9.6}, and we denote them $p_\mu^\pm\in H_\pm^3$, where $p_0^+>0$ and $p_0^- = - p_0^+ <0$. Antiparticles have no negative energy, because $E_-=-p_0^-=p_0^+=E_+>0$. We clarified this back in the non-relativistic case.

It follows from \eqref{10.3} that 
\begin{equation}\label{10.4}
p_\mu x^\mu =p_\mu x^\mu_0 + m\tau = m(\tau +\tau_0)\for m\tau_0:=p_\mu x^\mu_0\ ,
\end{equation}
and $\tau$ can be redefined such that $p_\mu x^\mu_0=0$ holds. For function $\mu_m=p^2/m$, the level surface $\mu_m=m$ is a disjoint union of 7-dimensional manifolds $H_\pm^3\times\R^{1,3}$. We denote the restriction of the 2-form \eqref{10.2} on them by
\begin{equation}\label{10.5}
\Omega_+=\dd p_\mu^+ \wedge \dd x^\mu =-\dd p_\mu^- \wedge \dd x^\mu =-\Omega_-\ ,\quad
p_\mu^\pm\in H_\pm^3\ .
\end{equation}
On 7-dimensional manifolds $H_\pm^3\times\R^{1,3}$ there acts a one-parameter translation group $\R_p$ with elements $g_\pm=\exp(\sfrac\tau{m}p^\mu_\pm\dpar_\mu )$ generating shifts \eqref{10.3}  along the trajectory. Identifying all points on the trajectories yields the space $X^6_{\mu_m}$. Therefore, the space $X^6_{\mu_m}$ parametrizing the trajectories \eqref{10.3} is the disjoint union of two cotangent bundles,
\begin{equation}\label{10.6}
T^*H^3 =T^*H^3_+\cup T^*H^3_-\ , \quad T^*H^3_\pm =\{p_\pm^2=m^2, p_\mu^\pm x_0^\mu =0, p_0^+=-p_0^->0\}\ .
\end{equation}
The restriction of the 2-forms \eqref{10.5} to $T^*H_\pm^3$ yields two non-degenerate symplectic forms on $T^*H_\pm^3$. 

\noindent{\bf Phase space $\C H^3$.} Particles and antiparticles are always related by a mapping of complex structures $\CJ\to-\CJ$ on manifold $X_{\mu_m}^6$ and complex bundles $L_\C^\pm$ over them. This is the charge conjugation operator $\hat C$ generated by the mapping $\tau\mapsto -\tau$ \cite{Popov3}. To illustrate this more clearly for the case of free particles \eqref{10.6}  under consideration, we identify $T^*H^3$ with the minimal complexification $\C H^3$ of the hyperboloid $H^3$.

Note that $T^*\R^{1,3}\cong\C^{1,3}$ is a complex space with coordinates $z^\mu = x^\mu -\im p^\mu$ and a pseudo-Hermitian metric of signature (1,3). We embed $T^*H^3$ into $\C^{1,3}$ as a complex quadratic hypersurface defined by the equation
\begin{equation}\label{10.7}
\C H^3=\left\{Z^\mu =\frac{m(x^\mu_0 -\im p^\mu )}{(p^2 + R^2)^{1/2}}  \with \eta_{\mu\nu}Z^\mu Z^\nu =-m^2    \right\}\ ,
\end{equation}
where
\begin{equation}\label{10.8}
R^2: = -\eta_{\mu\nu} x_0^\mu x_0^\nu >0\quad\mbox{since}\quad\eta^{\mu\nu}p_\mu p_\nu >0\und x_0^\mu\,\bot\, p_\mu\ .
\end{equation}
Both manifolds $T^*H^3_+$ (particles) and $T^*H^3_-$ (antiparticles) lie within $\C H^3$ and are related by an inversion of the complex structure $\CJ\to -\CJ$, which corresponds to the mappings $T^*H^3_+\to T^*H^3_-$ with $Z^\mu\mapsto\bar Z^\mu$, $p_\mu^+\mapsto p_\mu^-=-p_\mu^+$. Note that $\C H^3$ is the homogeneous space  SO(4, $\C$)$/$SO(3, $\C$) and is called the minimal complexification of the hyperboloid $H^3$. There also exists a non-minimal complexification $H_\C^3$ of the hyperboloid $H^3$, defined as the homogeneous space SU(1,3)$/$U(3), which describes the phase space of the Klein-Gordon oscillator \cite{Popov1, Popov2}. It is interesting that $H_\C^3$ is an open ball in $\C^3$ that also embeds into $\C H^3$; that is, the phase space of KG oscillator is a submanifold of the phase manifold of free KG particle.

\noindent{\bf Hilbert space $\CH_\pm$.}  We have described in detail the classical state spaces of free relativistic particles and antiparticles. Comparing this with the first step in \eqref{10.1}, we see that the phase space of relativistic particles remains six-dimensional; only its geometry has changed. Therefore, the second step in \eqref{10.1} -- the introduction of quantum  state spaces $\CH_\pm$ -- is essentially identical to the non-relativistic case.

On the manifold $\C H^3$, we define a holomorphic bundle $L_\C^+$ and antiholomorphic bundle $L_\C^-$, both Hermitian with structure group $\Uv$. Sections $\psi_+(Z)$ and $\psi_-(\bar Z)$ of these bundles have quantum charges (weights of group $\Uv$) equal to $q_+=1/\hbar$ and $q_-=-1/\hbar$. The scalar product on spaces of sections $\CH_\pm$ can be defined in the Bargmann-Segal approach using, for example, the Stenzel metric \cite{Stenzel, Szoke} or analytic continuation of the metric on $H^3$ to a tubular neighborhood in its complexification $\C H^3$  \cite{HallMit}. Returning from complex polarization to the momentum representation, we obtain an isomorphism between the Hilbert spaces $\CH_\pm$ and the spaces $L_\pm^2(H_\pm^3)$ of square-integrable functions $\phi_\pm(p_\pm)$ on hyperboloids $H_\pm^3$ with the measure dVol$_{H_\pm^3}^{}$. It should be emphasized that one cannot take the sum of functions  $\psi_+$ and $\psi_-$, or $\phi_+$ and $\phi_-$, without violating the polarization condition. Without this condition of representation irreducibility, we cannot treat them as wave functions.

\noindent{\bf Dynamics.} Having functions $\phi_\pm \in L_\pm^2(H_\pm^3)$, we proceed to the dynamics by following the sequence of steps \eqref{10.1}. Recall that in the non-relativistic case, the dynamics is determined by the evolution operators \eqref{7.7}. For free particles in momentum representation they have the form
\begin{equation}\label{10.9}
U_\pm(t) = e^{-\im q_\pm\frac{p^2}{2m}t}=e^{-\im q_\pm Et}\ ,
\end{equation}
where $q_\pm =\pm \sfrac1\hbar$. Note that here $E$ and $t$ are conjugate variables. In the relativistic case, the dynamics \eqref{10.4} is determined by the Casimir function $\sfrac{p^2}m$,
\begin{equation}\label{10.10}
U_\pm(\tau) = e^{-\im q_\pm\frac{p^2}{m}\tau}=e^{-\im q_\pm E_0\tau}\ ,\quad E_0=mc^2\ ,
\end{equation}
and this represents dynamics in terms of proper time $\tau$. It should be emphasized that $\tau$ is the variable conjugate to the particle's rest energy $E_0=mc^2$-- or mass, since the speed of light is a fundamental constant. This is precisely the meaning behind Wigner's classification of relativistic equations based on mass and spin.

Using formulae \eqref{10.4} and \eqref{10.6} in the form $mc^2\tau =p_\mu x^\mu$, we obtain an embedding of trajectories into space $H_\pm^3\times\R^{1,3}$, with
\begin{equation}\label{10.11}
U_\pm (\tau)\phi_\pm (p_\pm )= e^{-\frac\im\hbar p_\mu^\pm x^\mu}\phi_\pm (p_\pm)\ .
\end{equation}
On the right-hand side of \eqref{10.11}, we obtained  two plane waves with fixed momentum moving in Minkowski space, with $x^\mu\in\R^{1,3}$. Each of them satisfies the Klein-Gordon equation \eqref{9.4}.  It is important to understand here that $\phi_+$ cannot have negative-frequency modes and $\phi_-$ cannot have positive-frequency modes, by the very definition of these fields.
By integrating them over the spaces $H_\pm^3$, we obtain wave packets $\phi_\pm(x)$ from \eqref{9.5}. Recall, however, that $\phi_\pm$ are sections of different line bundles $L_\C^\pm$ and have opposite quantum charges $q_\pm =\pm\sfrac1\hbar$. Therefore, they cannot be added as in \eqref{9.5} since the sum ceases to be a wave function due to the loss of irreducibility. The irreducibility of the representation of the Poincare group and the group $\Uv$ is preserved for the functions $\phi_+$ and $\phi_-$ separately. Thus, we have derived the Klein-Gordon equation via the standard chain \eqref{10.1} of quantum mechanics and shown that they must be given separately for particles $\phi_+$ and antiparticles $\phi_-$, where the indices reflects their weights $\pm 1$ relative to the group $\Uv$.

\section{First-quantized Klein-Gordon theory}

\noindent{\bf Lagrangians and charges.} After we discussed in the previous section the derivation of the Klein-Gordon equations through the transition from classical to quantum state spaces, it is not difficult to reformulate the standard approach \eqref{9.1}-\eqref{9.5} into an approach that takes into account the quantum charges $q_\pm$ of the fields $\phi_\pm$.

The Schr\"odinger equation \eqref{7.8} for $\psi_+$ and $\psi_-$ are generalized to the relativistic case independently of each other. Therefore, one should consider two independent Lagrangians, 
\begin{equation}\label{11.1}
\CL_{\phi_\pm}^{}=\frac{\hbar^2}{2m}\left(\eta^{\mu\nu}\dpar_\mu\phi_\pm^*\dpar_\nu\phi_\pm - \Bigl(\frac{mc}\hbar\Bigr)^2\phi_\pm^*\phi_\pm\right)\ ,
\end{equation}
and take their sum only if we wish to have both states simultaneously. Unlike  \eqref{9.1},  these Lagrangians are invariant under the action of the group $\Uv$. From these Lagrangians follow the Klein-Gordon equations
\begin{equation}\label{11.2}
\left(\eta^{\mu\nu}\dpar_\mu\dpar_\nu + \Bigl(\frac{mc}\hbar\Bigr)^2\right)\phi_\pm=0
\end{equation}
for particles and antiparticles. The $\pm$ signs here are related with quantum charges $q_\pm$, not to the frequencies. These charges are linked to the right-hand side of the commutation relations \eqref{6.3} and the corresponding complex structure (holomorphicity/antiholomorphicity) on the phase space. In the case of free particles, this reduced to a correspondence with the signs of the frequencies.

From \eqref{11.1} follow two Hamiltonians
\begin{equation}\label{11.3}
\CH_{\phi_\pm}^{}=\frac{\hbar^2}{2m}\left(\delta^{\mu\nu}\dpar_\mu\phi_\pm^*\dpar_\nu\phi_\pm + \Bigl(\frac{mc}\hbar\Bigr)^2\phi_\pm^*\phi_\pm\right)
\end{equation}
and two density functions for quantum charges $q_\pm$,
\begin{equation}\label{11.4}
j_{\phi_\pm}^{t}=\frac{\im\hbar^2}{2mc^2}\left(\phi_\pm^*\dpar_t\phi_\pm - \phi_\pm\dpar_t\phi_\pm^*\right)\ .
\end{equation}
The conserved positive energy $E_+=E_-=E$ following from \eqref{11.3} is in accordance with the eigenvalues of the operators $\hat E_\pm=\pm\im\hbar\dpar_t$. The integrals of \eqref{11.4} over $\R^3$ yield time-conserved quantum charges $q_\pm$ and positive probabilities with densities $\pm j_{\phi_\pm}^{t}$. We emphasize that all Noether charges for the wave functions $\phi_+$ and $\phi_-$ are conserved independently.

\noindent{\bf Electric charges.} Physicists lack of understanding of what means for a field to possess an electric charge. The fields $\phi_\pm$ we have considered possess quantum charges $q_\pm$ of the group $\Uv$, but they lack electric charges, even though they are complex fields. Possessing charges $q_\pm$ implies an implicit dependence on $\exp (\im\theta )\in\Uv$ according to formula \eqref{4.5}. In purely algebraic terms, this is a statement about $\phi_\pm$ belonging to certain representation of the group $\Uv$. Similarly, if we wish to endow the field $\psi_\pm^{(n)}$ from \eqref{4.5} with electric charges $ke_\pm$ as well, we must assign them representation weights of the group U(1)$_{\sf em}$ and place the field $\psi_\pm^{(n,k)}$ in the tensor product of one-dimensional complex representations of these two groups -- $\Uv$ and 
U(1)$_{\sf em}$. That is precisely how we introduced charges $(nq_\pm , ke_\pm)$ with $n,k\ge 0$ in Section 7. 

Within the framework of relativistic Hamiltonian mechanics  \eqref{10.2} -\eqref{10.6}, the electromagnetic field $A^{\sf em}=A^{\sf em}_\mu\dd x^\mu$ is introduced via a deformation of the symplectic form \eqref{10.2}, considered for two values of $p_0$ as in \eqref{10.5} \cite{Sour}. For this generalization of the non-relativistic formulae \eqref{7.4}, we have
\begin{equation}\label{11.5}
\Omega_\pm^{\sf em}=\dd\Pi_\mu^\pm\wedge\dd x^\mu + e\dd A^{\sf em}=\dd \bigl[(\Pi_\mu^\pm + e A^{\sf em}_\mu)\dd x^\mu\bigr]=\dd p_\mu^\pm\wedge\dd x^\mu\ ,
\end{equation}
\begin{equation}\label{11.6}
\Rightarrow\Pi_\mu^\pm = p_\mu^\pm - e A^{\sf em}_\mu =\pm\im\hbar\dpar_\mu - e A^{\sf em}_\mu = \pm\im\hbar (\dpar _\mu \pm\frac{\im e}{\hbar}A_\mu^{\sf em})=:\pm\im\hbar D_\mu^\pm\ .
\end{equation}
It is evident from these formulae that, upon transitioning from the upper sheet of the hyperboloid \eqref{10.6} to the lower one, the sign of both the quantum charge and the electric charge changes. Thus, if we assign electric charges $e_\pm = \pm e$ to our fields $\phi_\pm$ (which already possess quantum charges $q_\pm$), they will interact not only with the vacuum field $\Av$ but also with the electromagnetic field $A_{\sf em}$  via the covariant derivatives introduced in \eqref{11.6}. In this case, the Lagrangians \eqref{11.1} are generalized as follows:
\begin{equation}\label{11.7}
\CL_{\phi_\pm}^{\sf em}=\frac{\hbar^2}{2m}\left[\eta^{\mu\nu}(D_\mu^\pm\phi_\pm)^*D_\nu^\pm\phi_\pm - \Bigl(\frac{mc}\hbar\Bigr)^2\phi_\pm^*\phi_\pm\right]\ .
\end{equation}
They lead to two independent equations
\begin{equation}\label{11.8}
\left(\eta^{\mu\nu}D_\mu^\pm D_\nu^\pm + \Bigl(\frac{mc}\hbar\Bigr)^2\right)\phi_\pm =0\ ,
\end{equation}
where the sign ``$\pm$" indicates that the fields $\phi_\pm$ carry charges $(q_\pm , e_\pm)$.

Note that from formulae \eqref{11.5}-\eqref{11.8} it follows that fields $\phi_+$ and $\phi_-$ have opposite charges. Therefore, the assumptions adopted in the standard approach -- that both $\phi_+$ and $\phi_-$ are particles with identical charges -- contradict the mathematics. These charges are assigned to the fields before the introduction of the Lagrangian. Moreover, the Lagrangian must be specified in a form invariant under the action of both group $\Uv$ and group U(1)$_{\sf em}$. In this case, the charges will be conserved quantities according to Noether's theorem. Therefore, the creation and annihilation of particle-antiparticle pair cannot occur in quantum mechanics. In an $\Uv$-invariant formulation of the theory, there is no Klein paradox. In quantum mechanics, the number of particles and the number of antiparticles are fixed independently, and these numbers cannot be altered by any change of external fields. This is possible only after second quantization, by specifying interaction terms that do not preserve $\Uv$-invariance. Quantum charges are not conserved in QFT, by the very construction of the theory.

\section{First-quantized Dirac theory}

\noindent{\bf Problems with the Dirac Lagrangian.} All the problems of the standard Dirac theory have the same causes as those of the standard Klein-Gordon theory. The Dirac equation for spinor $\psi\in\C^4$, 
\begin{equation}\label{12.1}
\bigl(\im\gamma^\mu\dpar_\mu - \frac{mc}\hbar\bigr)\psi =0\ ,
\end{equation}
has solutions of the same type $\psi =\psi_++\psi_-$ as the Klein-Gordon solutions \eqref{9.5}. The spinor-valued functions $\psi_+$ and $\psi_-$ are also parametrized by free functions on the manifolds
\begin{equation}\label{12.2}
T^*H_\pm^3\cong\C H_{\pm\CJ}^3\ ,
\end{equation}
where $\CJ$ is a complex structure on $T^*H_+^3$. These functions are polarized sections of the complex conjugate bundles $L_\C^+$ and $L_\C^-$ and have quantum charges $q_+=\sfrac1\hbar$ and $q_-=-\sfrac1\hbar$. The spinors $\psi_+$ and $\psi_-$ are the wave functions of particle and antiparticle, respectively.
Recall that the indices ``$\pm$" are weights of conjugate representations of the group $\Uv$, defined prior to the introduction of Lagrangians.

If the sum $\psi=\psi_++\psi_-$ of solutions $\psi_+$ and $\psi_-$ is taken, the polarization condition -- a necessary condition for defining the wave function -- will be violated. Accordingly, the Lagrangian from which equation \eqref{12.1} follows is not $\Uv$-invariant, analogous to what we discussed in \eqref{9.9} for the standard Klein-Gordon Lagrangian. Group $\Uv$ fixes the weights of the wave functions (quantum charges), that is, the number of particles and antiparticles, and these numbers are not conserved when $\Uv$-invariance is violated. All problems (including negative energies) disappear upon the introduction of $\Uv$-invariant Lagrangians defined so as to satisfy the correspondence principle with the non-relativistic Pauli equations for particles and antiparticles.

\noindent{\bf First-quantized theory.} Free particles and antiparticles of spin $s=\sfrac12$ are described by Lagrangians
\begin{equation}\label{12.3}
\CL_{\psi_{\pm}}=\pm\hbar c\bigl(\bar\psi_{\pm}\im\gamma^\mu\dpar_\mu\psi_\pm - \frac{mc}\hbar \bar\psi_{\pm}\psi_\pm\bigr)\ ,
\end{equation}
where $\bar\psi_{\pm}=\psi_\pm^\+\gamma^0$. From these Lagrangians follow the equations
\begin{equation}\label{12.4}
\bigl(\im\gamma^\mu\dpar_\mu - \frac{mc}\hbar \bigr)\psi_{\pm}=0\ ,
\end{equation}
where $\dpar_\mu =\dpar/\dpar x^\mu$. Let us emphasize once again that the indices ``$\pm$" refer to the quantum charges $q_\pm=\pm\sfrac1\hbar$ of independent fields $\psi_\pm$, rather than to frequencies with respect to the operator $\im\hbar\dpar_t$. The charges $q_\pm$ are defined in the general case where the frequency expansion may be absent altogether -- for example, for the Klein-Gordon oscillator \cite{Popov1, Popov2} or when moving in curved manifold in which Killing vectors are absent. The possession of charges $q_\pm$ is entirely independent of external fields. The charges $q_\pm$ are related to the orientation of the particle's world line, and charge conjugation $q_+\mapsto q_-$ is related to the mapping $\tau\mapsto -\tau$ of proper time.

Using Noether's theorem for time translation invariance, we obtain expressions  for the energy density for particles and antiparticles:
\begin{equation}\label{12.5}
\CH_{\psi_\pm}^{}=\psi_\pm^\+\left[\pm\hbar c (-\im\gamma^0\gamma^a\dpar_a +\frac{mc}\hbar\gamma^0)\right]\psi_\pm =
\psi_\pm^\+\hat H_D^\pm\psi_\pm\ .
\end{equation}
It is easy to see that for the plane-wave solutions
\begin{equation}\label{12.6}
\psi_+=e^{-\frac\im\hbar p_\mu^+x^\mu}u(p_+)\und\psi_-=e^{-\frac\im\hbar  p_\mu^-x^\mu}v(p_-)
\end{equation}
the energy densities are positive and equal, $\CH_{\psi_\pm}^{}=\pm p_0^\pm c= E$. Note that this is as it should be, since the energy operators for the particle and antiparticle sectors have the form $\hat E_\pm =\pm\im\hbar c\dpar_0$ ($=\hat H_D^\pm$ on solutions).

The invariance of the Lagrangians \eqref{12.3} under independent transformations $\psi_\pm\mapsto\exp(-\im\al_\pm)\psi_\pm$ of the group $\Uv$ implies the conservation over the time of the charges $q_\pm$ with densities
\begin{equation}\label{12.7}
j_\pm^t=\pm\psi_\pm^\+\psi_\pm\quad\Rightarrow\quad q_\pm=\frac1\hbar\int_{\R^3}j_\pm^t\dd^3x\ .
\end{equation}
The independent conservation of charges $q_+$ and $q_-$ make it possible to introduce probability densities
\begin{equation}\label{12.8}
\rho_+=\psi_+^\+\psi_+\und\rho_-=\psi_-^\+\psi_-
\end{equation}
for the detection of particles and antiparticles at a given point in space $\R^3$ at a fixed time $t$.

\noindent{\bf Non-relativistic limit.} The introduction of electric charges $e_\pm$ for quantum particles $\psi_\pm$ is described in detail in the preceding sections. We consider spinor-valued sections $\psi_\pm$ of bundles $L_\C^\pm\otimes E_{\sf em}^\pm$ having charges $(q_\pm , e_\pm)$. For such $\psi_\pm$, partial derivatives are everywhere replaced by covariant derivatives \eqref{11.6}, and instead of \eqref{12.4} we obtain the equations 
\begin{equation}\label{12.9}
\left (\im\gamma^\mu D_\mu^\pm - \frac{mc}\hbar\right)\psi_\pm =0\ .
\end{equation}
Adding fields $\psi_+$ and $\psi_-$ is possible only if they have the same charges, yet mathematics insists that they are opposite.
Note again that under no external field $A_\mu$ can the wave function $\psi_+$ be mapped into the wave function $\psi_-$, due to the conservation of the Noether charges $q_+,e_+$  and $q_-,e_-$. This is fully consistent with the Lorentz equations for charged classical particles. The covariant derivatives $D_\mu^\pm$ correspond to the kinetic momenta $\Pi_\mu^\pm$, and the particle's momentum lies on the upper sheet of the hyperboloid $\Pi_\pm^2=\eta^{\mu\nu}\Pi_\mu^\pm\Pi_\nu^\pm =m^2c^2$, while antiparticle's momentum lies on the lower sheet. 
 If we differentiate the functions $\Pi_\pm^2$ with respect to $\tau$ using the Lorentz equation, we get zero, that is, the particle's trajectory cannot move from the upper to the lower hyperboloid.

To pass to the non-relativistic limit $c\to\infty$, we set 
\begin{equation}\label{12.10}
\psi_\pm = e^{-\im q_\pm m c^2 t}\tilde\psi_\pm (x,t)\for \tilde\psi_+ =\begin{pmatrix}\chi_+\\\eta_+\end{pmatrix}\und
\tilde\psi_- =\begin{pmatrix}\eta_-\\\chi_-\end{pmatrix}\ ,
\end{equation}
analogously to formulae \eqref{9.7} and \eqref{9.8} for the Klein-Gordon equations. It is easy to verify that, upon substituting \eqref{12.10} into \eqref{12.9}, we obtain the Pauli equations \eqref{8.5} for $\chi_\pm$ in the non-relativistic limit. In this case, the fields $\eta_\pm$ are expressed in terms of $\chi_\pm$ by the formulae:
\begin{equation}\label{12.11}
\eta_\pm=\mp\frac{\im\hbar}{2mc}\sigma^aD_a^\pm\chi_\pm +\CO\left(\frac1{c^2}\right)\ .
\end{equation}
Thus, for particles of spin $s=\sfrac12$ we have a complete correspondence between the relativistic Dirac equations and the non-relativistic Pauli equations for particles and antiparticles. The same holds for the two Klein-Gordon and the two Schr\"odinger equations, which is in full agreement with the correspondence principle. The creation and annihilation of particles become possible only after the transition to quantum field theory.

\bigskip

\noindent 
{\bf\large Acknowledgments}

\noindent
I am grateful to Tatiana Ivanova for useful remarks.

\bigskip


\end{document}